# Consistent estimation of the critical current density of a superconductor from hysteresis loops

Ratan Lal

Superconductivity Division, National Physical Laboratory, Dr. K. S. Krishnan Road,

New Delhi, India

## Abstract

It has been noticed that the critical current density $J_c$ of some of the superconducting samples, calculated on the basis of Bean model, increases with increasing magnetic field $H$ up to a significant range above $H$=0. This is an inconsistent behavior of $J_c$ since the theory of Kim and the theories based on vortex dynamics, all, lead to decreasing $J_c$ with increasing $H$ for $H > 0$. It has been argued that a realistic variation of $J_c$ for low $H$ may be obtained within Bean framework by redefining the width of the hysteresis loop. The new definition of the loop width is guided by the requirement that $J_c$ stays as close to the $J_c$ of the theory of Kim as possible. Illustrative calculations of $J_c$ show its considerable enhancement over the Bean values.

## 1. Introduction

In a recent publication Cheng et al have presented values of the critical current density ($J_c$) of various sample of the $MgB_2$ superconductor [1]. The $J_c$ of one of these samples, namely that with 8% nano diamond (ND), decreases at 10 K with decreasing magnetic field ($H$) below about 0.5 T. This is an unexpected behavior of the $H$-dependence of $J_c$. In order to see why it is so, we first consider the lower critical field ($H_{c1}$) of the $MgB_2$ superconductor. By using the values of the penetration depth, $\lambda$=140 nm [2], and coherence length, $\xi$=5.2 nm [2] in Eq. 3.56 of de Gennes [3] we find that $H_{c1}$ turns out to be about 0.03 T, which is practically zero on the scale of the irreversibility field ($H_{irr} \sim 6 - 10\ T$ [1]). Thus, as soon as the magnetic field is applied to the superconductor from $H = 0$, vortices will start to enter in the system. The movement of these vortices will increase with increasing magnetic field [4] so that $J_c$ will decrease with increasing $H$. This will happen at least for low $H$ irrespective of the model used for describing the vortex dynamics (cf. Eqs. 6, 7, 13, 16, 25, 31 and 40 of Wordenweber [4]). In fact, in general, this will continue up to higher $H$ also, but for the case like that of collective pinning [5] $J_c$ will increase for a portion of $H$ near $H_{irr}$. From the work of Chen and Goldfarb [6] it becomes clear that in the Kim's theory also $J_c$ decreases with increasing $H$.

Vajpayee et al [7] have also made a study of $J_c$ of ND-doped $MgB_2$ samples. The 5% ND and 7% ND samples of these authors show increasing $J_c$ with $H$ at 10 K below about 1.5 T. In fact, this behavior of $J_c$ has been encountered earlier also, for example, by Niu and Hampshire [8] in the sample numbers 6 and 7 of $PbMo_6S_8$ below about 0.5 T. (On the basis of Ref. [8] it may be shown that $H_{c1}$ of $PbMo_6S_8$ will be of order of 0.0002 T, while $H_{irr}$ is of order of 25 T.) Many more references may be found out in literature where $J_c$ increases with $H$ in the low-$H$ regime. However, for specificity, we shall limit to the references [1], [7] and [8] only.

The above-mentioned increasing $J_c$ with increasing $H$ in the low magnetic field regime arises due to the Bean's formulation [9] since all of the above authors have used this method for estimating $J_c$ from the *M-H* curves. From Fig. 8 of Chen and Goldfarb it becomes clear that the Bean's $J_c$ shows increasingly larger deviation from the Kim's $J_c$ near $H$ =0 when the magnetic moment corresponds to increasingly stronger dependence on *H*. Since the Bean's theory is a special case of the Kim's theory [6], the latter is more realistic than the Bean's theory. In this sense a deviation of the Bean's and Kim's $J_c$ values for low $H$ implies inadequacy of the Bean's theory for this region of low $H$. Thus there is a need for a realistic method for estimating $J_c$ from hysteresis loops. It may be noted that knowledge of the realistic $J_c$ is important not only from the view-point of its magnitude, but also from the view-point of pinning mechanism. This is because the pinning mechanism is understood by looking at the variation of the pinning force density $F_p = \mu_0 H J_c(H) \propto (\frac{H}{H_{irr}})^p (1 - \frac{H}{H_{irr}})^q$ with $H$ [1]. (Here $\mu_0$ is free space permeability.) If $J_c(H)$ is not realistic, the values of $p$ and $q$ may be unrealistic too, leading to a misleading interpretation of the vortex dynamics. Thus a realistic way for estimating $J_c$ from hysteresis loops is highly desired.

This task has been performed in the present article. We suggest a method for extracting values of the critical current density by redefining the width of the hysteresis loop at a particular $H$ such that the resulting $J_c$ stays as close to the Kim's $J_c$ as possible.

## 2. Formalism

Let $M^+(H)$ and $M^-(H)$ denote respectively the positive and negative parts of the magnetic moment of a hysteresis loop. Then, according to Bean's formulation, $J_c$ is given by [6]

$$J_c(H) = G[M^+(H) - M^-(H)]. \tag{1}$$

Here $G$ is a geometric factor.

In order to see why Bean's formula results in an inconsistent behavior of $J_c$ for low $H$, we proceed as follows. In the Bean's theory the critical current density is considered to be independent of the magnetic field [6]. In the sense of Eq. (1) this means

$$\frac{dM^+(H)}{dH} \approx 0; \qquad \frac{dM^-(H)}{dH} \approx 0 \tag{2}$$

On the other hand, the magnetic moment of $MgB_2$ and $PbMo_6S_8$ superconductors changes rapidly with $H$. That is to say,

$$\frac{dM^+(H)}{dH} \gg 0; \qquad \frac{dM^-(H)}{dH} \gg 0 \tag{3}$$

This can be seen, for example, from Fig. 2 of Vajpayee et al [7] and from the inset of Fig. 9 of Niu and Hampshire [8].

The situation of Eq. (3) enhances the possibility of the variation of the right-hand side of Eq. (1) with $H$ to be positive. That is to say, for

$$\frac{dM^+(H)}{dH} - \frac{dM^-(H)}{dH} > 0 \tag{4}$$

to be satisfied for low $H$. When this occurs, Eq. (1) will lead to a $J_c$ which increases with $H$ for low $H$. Thus, the inadequacy of the Bean's formalism at low $H$ arises due to fast variation of the moments $M^+(H)$ and $M^-(H)$ with $H$.

In order to clarify Eq. (4) we consider the positions of the maximum (minimum) of the moment $M^+$ ($M^-$). The maximum of $M^+(H)$ lies in the negative-$H$ side, while that of $M^-(H)$ lies on the positive-$H$ side such that both are equidistance from $H$=0. (cf. Fig. 6e of Chen and Goldfarb.) In this sense let $-H_{max}$ ($H_{max} > 0$) be the position of the maximum of $M^+(H)$, then $H_{max}$ will be the position of the minimum of $M^-(H)$. If $H_{max} = 0$, condition of Eq. (4) will never be satisfied because the left-hand side of this equation will be essentially negative. In fact, for $H_{max} = 0$, $\frac{dM^+(H)}{dH}$ will be negative, while $\frac{dM^-(H)}{dH}$ will be positive. When $H_{max}$ increases beyond 0, $\frac{dM^+(H)}{dH}$ will remain negative, but $\frac{dM^-(H)}{dH}$ will change sign from positive to negative between $H$=0 and $H$= $H_{max}$. Thus if $H_{max}$ is sufficiently away from $H$=0, a situation will arise when $|\frac{dM^-(H)}{dH}|$ will become larger than $|\frac{dM^+(H)}{dH}|$. When it happens so, Eq. (4) will be satisfied and, according to Eq. (1), $J_c$ will increase with $H$ up to $H = H_{max}$.

Thus it is the value of $H_{max}$ which leads to values of $J_c$ as found by Cheng et al [1], Vajpayee et al [7], and Niu and Hampshire [8] for low $H$ for some of the superconducting samples. The value of $H_{max}$ is 0.083$H_{irr}$ for the Fig. 6e of Chen and Goldfarb [6]. Cheng et al [1], Vajpayee et al [7], and Niu and Hampshire [8] have not given hysteresis loops for the above-mentioned samples. So, it is difficult to estimate accurate vales of $H_{max}$ for these samples. However, on the basis of the variation of $J_c$ of these samples a rough estimate can be made. We find $H_{max}$= 0.5 T, 1.5 T, 1.5 T, 0.5 T and 0.5 T respectively for the 8%ND sample [1], 5% ND sample [7], 7%ND sample [7], sample number 6 [8] and sample number 7 [8].

Let us see what the Kim's theory, of which Bean's theory is a special case [6], say about the behavior of $J_c$ for the situation of Eq. (3). Looking at the various parts of Fig. 6 of Chen and Goldfarb [6] we find that the condition of Eq. (3) is most satisfied for Fig. 6e. Fig. 8e shows the values of $J_c$ corresponding to this figure. We see that there is perfect agreement between the Bean's $J_c$ and Kim's $J_c$ for $H > H_p$, where $H_p$ is full penetration field. Below $H_p$ the difference between these two theories become increasingly larger with decreasing $H$. While the Bean's $J_c$ changes curvature below $H_p$ so that it bends downward near $H$=0, the Kim's $J_c$ continues the same curvature down to $H$=0 tending to infinity for $H$=0. In fact, for the situation of Fig. 8e Kim's $J_c$ behaves like $1/H$.

The above comparison of the Bean's $J_c$ and Kim's $J_c$ makes it clear that for the situation of Eq. (3) these two critical currents move in opposite directions. Because of the change in curvature the Bean's $J_c$ becomes lower than the realistic $J_c$ near $H$=0. On the other hand, because of the fact that $J_c \rightarrow \infty$ for $H \rightarrow 0$, the Kim's $J_c$ will be larger than the realistic $J_c$ near $H$=0. So the realistic $J_c$ will lie in between the Bean's $J_c$ and Kim's $J_c$. Below we describe a method to estimate this realistic $J_c$.

We take the magnetic moments $M^+(H)$ and $M^-(H)$ as input, but redefine the width of the hysteresis loop,$\Delta M(H)$, at the magnetic field $H$. For this purpose we, first of all, note that for $H > H_p$ Bean's $J_c$ and Kim's $J_c$ lead to the same set of values [6]. So, we take

$$\Delta M\big(H > H_p\big) = M^+(H) - M^-(H). \tag{5}$$

We now consider the $H$=0 point. Since the sought-for $J_c$ is required to have positive curvature for all $H$, it will be maximum at $H$=0. Moreover, we require that the sought-for $J_c$ remains as close to the Kim's $J_c$ as possible. The maximum possible value of $\Delta M(H=0)$ from the moments $M^+(H)$ and $M^-(H)$ is given by

$$\Delta M(H=0) = M^+(-H_{max}) - M^-(+H_{max}). \qquad (6)$$

We are now left with the values of $\Delta M(H)$ for $0 < H < H_p$. For this purpose we compress the values of $M^+(H)$ from a range of width $H_{max} + H_p$ ($-H_{max} \leq H \leq H_p$) to a shorter range of width $H_p$ ($0 \leq H \leq H_p$). Such a task is performed in a practically convenient way by the function

$$s(H) = \exp\left(-\frac{uH}{H_p}\right) \qquad (7)$$

such that

$$e^{-u} \approx 0. \qquad (8)$$

From Eq. (8) we can see that s(0)=1 and $s(H > H_p) \approx 0$. Using the function s($H$) we can stretch the moment values $M^-$ from the shorter range of width $H_p$-$H_{max}$ ($H_{max} \leq H \leq H_p$) to a range of width $H_p$ ($0 \leq H \leq H_p$). The loop width $\Delta M(H)$ for $0 < H < H_p$ is now expressed in terms of the compressed moment $M^+$ and stretched moment $M^-$ as given by

$$\Delta M(H) = M^+(H - H_{max}s) - M^-(H + H_{max}s). \qquad (9)$$

This equation tends to Eq. (6) for $H$=0, and to Eq. (5) for $H \gg H_p$. We replace the Bean's loop width by this new width so that Eq. (1) for the critical current density is modified to

$$J_c(H) = G\Delta M(H). \qquad (10)$$

For $H_{max} \approx 0$ this equation tends to the Bean's formula (Eq. 1).

## 3. Results and discussion

In order to clarify the importance of Eqs. (9) and (10) we have calculated $J_c(H)$ using the hysteresis loop of Fig. 6e of Chen and Goldfarb [6]. The results are shown in the third column of Table 1 for various values of $H/H_p$. In the calculations we have taken u=7, which guarantees that Eq. (5) gets satisfied to within an error of 0.001. The second (fourth) column of this table corresponds to the Bean's (Kim's) $J_c$ read from Fig. 8e of Chen and Goldfarb. From table 1 we see that the present values of $J_c$ matches with that of Kim's $J_c$ down to $H = 0.6H_p$, while that of the Bean's $J_c$ matches with the Kim's $J_c$ down to $H = H_p$ only. This shows that the present method leads to indeed more realistic $J_c$ than the Bean's $J_c$. The deviation of the present $J_c$ from Kim's $J_c$ below $H = 0.6H_p$ occurs because the latter diverges at $H$=0, while the present $J_c$ is limited to a finite value by the width of the hysteresis loop.

The value of $J_c$ obtained by using Eqs. (9) and (10) for $H$=0 is $4.40 J_c(H_p)$ (cf. Table 1). This is significantly larger than the corresponding Bean's value, $3.21 J_c(H_p)$. Apart from this quantitative difference, the main difference lies in the qualitative sense. While the present $J_c$ continues increasing for decreasing $H$ down to $H=0$, the Bean's $J_c$ changes curvature at $H = 0.6H_p$ (cf. Fig. 6e of Ref. [6]). We emphasize that this change of curvature in the Bean's method is responsible for lower $J_c$ values near $H=0$. The value of $H_{max}$ for Fig. 6e of Chen and Goldfarb, as mentioned above, is $0.083H_{irr}$. If the value of $H_{max}$ increases further then after some stage we expect that Eq. (4) gets satisfied. When it happens so, $J_c$ will decrease for decreasing $H$ near $H=0$ in the Beans model, but according to the present estimation (Eqs. 9 and 10) $J_c$ will continue increasing for decreasing $H$. An important result of this illustration is that Bean's $J_c$ will need modification if $H_{max} >0$, irrespective of how $J_c$ varies near $H=0$. Moreover, near $H=0$ the present method will be more realistic than the Kim's method also because the latter gives diverging $J_c$ at $H=0$. In fact, a superconductor can never support an infinite $J_c$. The upper limit of $J_c$ will be $J_{c,max} = n^* e\hbar/m\xi$ where $n^*$ is superfluid density, $e$ is magnitude of electron's charge, $\hbar$ is reduced Planck's constant, and $m$ is electron's mass. For the $MgB_2$ superconductor $J_{c,max}$ will be of the order of $10^8$ A/cm$^2$.

Hysteresis loops are not available in the articles of Cheng et al [1], Vajpayee et al [7] and Niu and Hampshire [8]. So, we are not in a position to present values like those in table 1 for the samples considered in these references. However, from the variation of $J_c$ we can get rough idea about $H_{max}$ and variations of the magnetic moments with $H$ near $H=0$. Such values of $H_{max}$ and $J_c(H)$ are given in table 2. It is clear from this table that the present method modifies $J_c(H)$ considerably.

It may be noted that the enhancement of $J_c(H)$ in the present method will shift the critical force density $F_p$ towards $H = 0$. This will lower the peak position, $p/(p+q)$, of $F_p$ versus $H/H_{irr}$ curve, thereby affecting the nature of pinning mechanism.

## 4. Conclusions

In the present paper we have pointed out cases where the critical current density of some samples of MgB$_2$ and PbMo$_6$S$_8$ increases with magnetic field for low $H$. Since in these systems the values of the lower critical field are practically zero on the scale of the irreversible field, this trend of $J_c(H)$ is unrealistic. We have identified the origin of this behavior of $J_c(H)$ for low $H$ in the sharp variation of the magnetic moment with $H$ (Eq. 3) combined with significantly larger values of $H_{max}$. For a consistent extraction of the critical current density from the hysteresis loops we have suggested a method, Eq. (9) and (10), which is governed by the condition that the new values of $J_c(H)$ stay as close to the Kim's values as possible. Although the present method is motivated by increasing $J_c(H)$ with $H$, it has a larger range of applicability in that it is suitable for any superconductor satisfying Eq. (3) and having sufficient value of $H_{max}$ irrespective of whether $J_c(H)$ increases or decreases with $H$ for low $H$.

**Table 1**: Values of the relative critical current density $J_c(H)/J_c(H_p)$ for various values of $H/H_p$ in the present case, Bean's formalism and Kim's theory. The latter two sets of values are read from the Fig. 8e of Chen and Goldfarb [6], while the present values are obtained on the basis of the hysteresis loop of Fig. 6e of Ref. [6].

| $H/H_p$ | $J_c(H)/J_c(H_p)$ | | |
|---------|------|---------|-----|
|         | Bean | Present | Kim |
| 0.0 | 3.21 | 4.40 | $\infty$ |
| 0.2 | 3.13 | 3.13 | 5.18 |
| 0.4 | 2.90 | 2.23 | 2.45 |
| 0.6 | 2.18 | 1.67 | 1.67 |
| 0.8 | 1.45 | 1.18 | 1.18 |
| 1.0 | 1.00 | 1.00 | 1.00 |

**Table 2**: Values of $H_{max}$ and $J_c(H)$ for different superconducting samples. The Bean's values are read from the respective references.

| Sample | Ref. | $\mu_0 H_{max}$ (T) | $\mu_0 H$ (T) | $J_c(H)$ $(10^5$ A/cm$^2)$ | |
|--------|------|------|------|------|------|
| | | | | Bean | Present |
| 8%ND MgB$_2$ | 1 | 0.5 | 0.0 | 3.98 | 5.75 |
| 5%ND MgB$_2$ | 7 | 1.5 | 1.0 | 2.01 | 3.19 |
| 7%ND MgB$_2$ | 7 | 1.5 | 1.0 | 0.60 | 1.59 |
| No. 6 PbMo$_6$S$_8$ | 8 | 0.5 | 0.0 | 2.03 | 2.54 |
| No. 7 PbMo$_6$S$_8$ | 8 | 0.5 | 0.0 | 1.04 | 1.30 |